\documentstyle[12pt]{article}

\begin{document}
{\sf \begin{center} \noindent
{\Large \bf A Riemannian geometrical method to classify tearing instabilities in plasmas}\\[3mm]

by \\[0.3cm]

{\sl L.C. Garcia de Andrade}\\

\vspace{0.5cm} Departamento de F\'{\i}sica
Te\'orica -- IF -- Universidade do Estado do Rio de Janeiro-UERJ\\[-3mm]
Rua S\~ao Francisco Xavier, 524\\[-3mm]
Cep 20550-003, Maracan\~a, Rio de Janeiro, RJ, Brasil\\[-3mm]
Electronic mail address: garcia@dft.if.uerj.br\\[-3mm]
\vspace{2cm} {\bf Abstract}
\end{center}
\paragraph*{}
Riemannian geometrical tools, such as Ricci collineations and
Killing symmetries , so often used in Einstein´s general theory of
gravitation are here applied to plasma physics to build magnetic
surfaces from Einstein plasma metrics used in tokamak devices. It is
shown that the Killing symmetries are constrains the Einstein
magnetic surfaces while the Killing vectors are built in terms of
the displacement of the toroidal surface. The pressure is computed
by applying these constraints to the pressure equations in tokamaks.
A method, based on the sign of the only nontrivial constant Riemann
curvature component, is suggested to classify tearing instability.
Throughout the computations two approximations are considered: The
first is the small toroidality and the other is the small
displacement of the magnetic surfaces as Einstein spaces.
\vspace{0.5cm} \noindent {\bf PACS numbers:}
\hfill\parbox[t]{13.5cm}{02.40.Hw-Riemannian geometries}

\newpage
\section{Introduction}
 The tools of the Riemann geometry \cite{1} so often used in other important areas of physics, such as Einstein theory of gravitation,
 have also been used by Mikhailovskii \cite{2} to investigate the tearing and other sort of
 instabilities in confined plasmas \cite{3}, where the Riemann metric tensor played a dynamical role interacting with the magnetic field
 through the magnetohydrodynamical equations (MHD). More recentely Garcia de Andrade \cite{4} has applied Riemann metric
 to investigate magnetic flux tubes in superconducting plasmas. Earlier Thiffault and Boozer \cite{5} have investigated the Riemann
 geometry in the context of chaotic flows and fast dynamos. In this paper we use the tools of Riemannian geometry,
 also used in other branches of physics as general relativity \cite{6}, such as Killing symmetries , Riemann and Ricci
 \cite{7} collineations , shall be applied here to generate magnetic nested surfaces in Tokamaks through the built of Einstein
 spaces obtained from Tokamak plasma metric. This work is motivated by the fact that the magnetic surfaces are severely
 constrained in tokamaks \cite{3} and Killing symmetries are well applied every time we have symmetries in the problem as in
 solutions of Einstein equations of general relativity. Equilibrium of these surfaces or their instabilities are fundamental
 in the constructio of tokamaks and other plasma devices such as stellarators where torsion is also present. The magnetic surfaces
 are more easily obtained when symmetries are present. This is our main motivation to apply the special Riemann geometrical techniques
 of Killing symmetries and Ricci collineation to obtain magnetic surfaces formed by Einstein spaces. To simplify matters we shall consider
 two usual approximations from plasma physics \cite{3} which are the small toroidality or inverse aspect ratio ${\epsilon}=\frac{a}{R}<<1$ where here R represents the external
 radius of the torus and a is its internal radius, and the Shafranov displacement ${\Delta}<<1$ as well. The pressure of
 the tokamak is obtained from the tokamak Shafranov shift equation. Constant pressure closed ergodic nested surfaces in magnetohydrostatics  have also been shown
 by Schief \cite{8} to be generated by solitons. This is another mathematical technique , distinct from ours, is another use of
 mathematical theory to generate of nested magnetic surfaces in plasmas. The paper is organised as follows:
 In section 2 we review the Riemannian technique of Killing vector and Ricci and Riemann colineations not usually familiar
 to the plasma physicists and solve the Ricci tensor components from the plasma metric by considering
 that nested surfaces are formed by Einstein spaces, where the Ricci tensor is proportional to the metric, and solve the
 Ricci collineation equations to find out the Killing vectors stablishing a geometrical method for the classification of tearing instabilities.
 Conclusions are presented in section 3.
 \section{Ricci collineations from plasma metrics and tearing instabilities}
 dynamos. Let us now start by considering the plasma metric given by Zakharov and Shafranov \cite{9} to investigate the evolution of
 equilibrium of toroidal plasmas. The components $g_{ik}$
 (i,k=1,2,3) and $(a,{\theta},z)$ as coordinates, of their plasma metric are
\begin{equation}
g_{11}=1-2{\Delta}'cos{\theta}+{{\Delta}'}^{2} \label{1}
\end{equation}
\begin{equation}
g_{22}=a^{2} \label{2}
\end{equation}
\begin{equation}
g_{33}=(R-{\Delta}+acos{\theta})^{2} \label{3}
\end{equation}
\begin{equation}
g_{12}=a{\Delta}'sin{\theta} \label{4}
\end{equation}
where $z=asin{\theta}$ and the dash represents derivation with
respect to a. In our approximation the last term in the expression
(\ref{1}) may be dropped. The magnetic surface equations are tori
with circular cross-section and equations
\begin{equation}r= R-{\Delta}(a)+acos{\theta} \label{5}
\end{equation}
\begin{equation}
z=asin{\theta} \label{6}
\end{equation}
Let us now compute the Riemann space of constant curvature
represented by the Riemann tensor components
\begin{equation} R_{ijkl}={\Lambda}(g_{ik}g_{jl}-g_{il}g_{jk}) \label{7}
\end{equation}
where ${\Lambda}$ ia constant which is called de Sitter cosmological
constant. Contraction of expression (\ref{7}) in two non-consecutive
indices,otherwise the symmetry of the Riemann curvature tensor
$R_{ijkl}= -R_{jikl}=R_{jilk}$ would make them vanish, yields the
Einstein space Ricci relation
\begin{equation}
R_{ik}=2{\Lambda}g_{ik}\label{8}
\end{equation}
 The Ricci collineations equations are given by
\begin{equation}
[{\partial}_{l}R_{ik}]{\eta}^{l}+R_{il}{\partial}_{k}{\eta}^{l}+R_{kl}{\partial}_{i}{\eta}^{l}=0
\label{9}
\end{equation}
where ${\eta}^{l}$ are the components of the Killing vector
$\vec{\eta}$ which defines the symmetries of the associated space,
and ${\partial}_{l}:=\frac{{\partial}}{{\partial}x^{l}}$ are the
components of the partial derivative operator. This equation is
obtained from the more elegant definition in terms of the Lie
derivative ${\cal L}_{\eta}$ as
\begin{equation}
{\cal L}_{\eta}R_{ik}=0 \label{10}
\end{equation}
In the next section we shall solve the Ricci collineation equations
in terms of the plasma metric above. \section{Nested surfaces as
Einstein spaces in plasmas} Let us now consider the application of
the above plasma metric into the Ricci collineation equation, which
yields
\begin{equation}
[{\partial}_{l}R_{11}]{\eta}^{l}+2R_{1l}{\partial}_{1}{\eta}^{l}=0
\label{11}
\end{equation}
\begin{equation}
[{\partial}_{l}R_{22}]{\eta}^{l}+2R_{2l}{\partial}_{2}{\eta}^{l}=0
\label{12}
\end{equation}
 We shall consider now just two independent coordinates
$(x^{1}=a,x^{2}={\theta})$ since nested surfaces are bidimensional
in the case of plasmas, Let us now compute the Riemann tensor
components in the linear approximation
\begin{equation}
R_{1212}=+\frac{{\partial}^{2}g_{12}}{{\partial}x^{1}{\partial}x^{2}}-\frac{{\partial}^{2}g_{11}}{{\partial}x^{2}{\partial}x^{2}}
-\frac{{\partial}^{2}g_{22}}{{\partial}x^{1}{\partial}x^{1}}
\label{13}
\end{equation}
Substitution of the plasma metric above into expression (\ref{13})
yields the expression
\begin{equation}
R_{1212}=[{2}+(asin{\theta}+2cos{\theta})] \label{14}
\end{equation}
It is easy to show that the components $R_{1313}$ and $R_{2323}$
both vanishes within our approximations.  At this point we consider
that ${\theta}$ is so small that $sin{\theta}$ vanishes and
$cos{\theta}=1$ this simplifies extremely our metric and turns it
into a diagonal metric where $g_{12}=0$ and $g^{bb}=(g_{bb})^{-1}$
$(b=1,2)$ and this allows us to compute the components of the Ricci
tensor from the Riemann component. But before that let us compute
the use the condition that the nested surface is an Einstein space
to compute the Riemann component again
\begin{equation}
R_{1212}={\Lambda}a^{2}[1-2{\Delta}'cos{\theta}] \label{15}
\end{equation}
Since both expressions for the Riemann component $R_{1212}$ must
coincide, equating expressions (\ref{15}) and (\ref{14}) yields an
expression for the derivative of the Shafranov shift ${\Delta}$ as
\begin{equation}
{\Delta}'=[1+\frac{{\Lambda}a^{2}}{2}] \label{16}
\end{equation}
Integration of this expression yields the value of the shift in
terms of the radius a as
\begin{equation}
{\Delta}=[a+\frac{{\Lambda}a^{3}}{6}] \label{17}
\end{equation}
Since the radius a is assumed to be small we may neglect terms of
the order $a^{3}$ which yields
\begin{equation}
{\Delta}=a \label{18}
\end{equation}
which satisfies the well-known boundary condition ${\Delta}(0)=0$.
From these expressions one may also compute ${\Delta}"={\Lambda}a$.
An important result in plasma physics is that tearing instabilities
coming from ion or electron currents possess the shift condition
${\Delta}'<0$. This condition would be clearly violated from
expression (\ref{16}) unless the ${\Lambda}$ curvature constant
would be negative and in modulus $|\frac{{\Lambda}a^{2}}{2}|>1$.
Note that this situation is very similar to the condition of
favorable or unfavorable curvature for the instabilities in plasmas
\cite{3}. The main difference is that here we are referreing to
Riemann curvature and not to Frenet curvature of the magnetic lines
in plasmas. This suggests another method to classify geometrically
tearing instabilities. Actually, since has been shown \cite{11}
recently that the Riemann tensor in plasmas can be expressed in
terms of the Frenet curvature both methods seems to be equivalent.
Now let us compute the Ricci components $R_{11}$ and $R_{22}$ from
the component $R_{1212}$ by tensor contraction with metric
components $g^{11}$ and $g^{22}$, which results in the expressions
\begin{equation}
R_{11}=-\frac{2}{a^{2}}[1+{\Delta}'] \label{19}
\end{equation}
and
\begin{equation}
R_{22}=[2+6{\Delta}']\label{20}
\end{equation}
which in turn yields the expressions
\begin{equation}
{\partial}_{1}R_{11}=-\frac{4}{a^{3}}[1+{\Delta}']
-\frac{2}{a^{2}}[{\Delta}"]=\frac{8}{a^{3}}\label{21}
\end{equation}
and
\begin{equation}
{\partial}_{2}R_{11}=-\frac{{\Delta}'}{a}=-\frac{(1+\frac{{\Lambda}{a}^{2}}{2})}{a}\label{22}
\end{equation}
\begin{equation}
{\partial}_{1}R_{22}={\Delta}'\label{23}
\end{equation}
\begin{equation}
{\partial}_{2}R_{22}=a{\Delta}'\label{24}
\end{equation}
Substitution of these derivatives of the Ricci tensor components
into the Ricci collineations equations one obtains the following set
of PDE equations
\begin{equation}
[a^{-1}+\frac{\Lambda}{4}]{\partial}_{1}{\eta}^{1}-{\eta}^{1}\frac{1}{a^{3}}=0
\label{25}
\end{equation}
which yields
\begin{equation}
{\eta}^{a}=exp[\frac{1}{a}] \label{26}
\end{equation}
\begin{equation}
\frac{16}{a}{\partial}_{2}{\eta}^{2}+{\eta}^{2}+6{\Lambda}{a}exp[\frac{1}{a}]=0
\label{27}
\end{equation}
which by considering the gauge ${\partial}_{2}{\eta}^{2}=0$ one
obtains the constraint
\begin{equation}
{\eta}^{\theta}=-6{\Lambda}exp[\frac{1}{a}]\label{28}
\end{equation}
for the Killing vector field poloidal component. The relation
\begin{equation}
\frac{{\eta}^{a}}{{\eta}^{\theta}}=-\frac{1}{6}{\Lambda}^{-1}\label{29}
\end{equation}
allows us finally to write down the Killing vector as
\begin{equation}
\vec{\eta}=-\frac{1}{6}{\Lambda}^{-1}[1,
-6{\Lambda}]{\eta}^{\theta}\label{30}
\end{equation}
which depends again upon the curvature scalar ${\Lambda}$ which is
central in the classification of tearing instabilities.
\section{Conclusions}
 In conclusion, we have investigated a method of classification and identification of tearing instability,
 allowing for example to distinguish between tearing instabilities that comes from ions and electron currents or
 not, based on the Riemann curvature constant submanifolds as nested
 surfaces in Einstein spaces. The Killing symmetries are shown also
 to be very useful in the classification of plasma metrics in the
 same way they were useful in classifying general relativistic
 solutions of Einstein's gravitational equations in four-dimensional
 spacetime \cite{6}. Other interesting examples of the utility of
 th is method is the Ricci collineations investigations of the
 twisted magnetic flux tubes and the Arnolds metric for the fast
 dynamo \cite{10,11,12}.
 \section*{Acknowledgements}
 Thanks are due to CNPq and UERJ for financial supports.


\newpage
\begin{thebibliography}{12}
\bibitem{1} E. Cartan, Riemannian geometry in an orthonormal Frame,
(2001) Princeton University Press.
\bibitem{2} A. Mikhailovskii, Instabilities in a Confined Plasma,
(1998) IOP.
\bibitem{3} R. White, The theory of toroidally confined
Plasmas,revised second edtion (2006) Imperial College Press.
\bibitem{4} L. C. Garcia de Andrade, Curvature and Torsion effects
on carrying currents twisted solar loops, (2006) Phys of Plasmas nov
issue.
\bibitem{5} J. Thiffault and A.H.Boozer, The Onset of Dissipation in the Kinematic Dynamo,Los Alamos arXiv:nlin.CD/0209042v1.L.C. Garcia de Andrade, Physics of Plasmas 13, 022309 (2006).
\bibitem{6} R. Penrose and W. Rindler, Spinors and spacetime vol1,
Oxford University Press (1984).
\bibitem{7} G. Ricci, Tensor Analysis,Boston.
\bibitem{8} W.K. Schief, J. Plasma Physics 69 (2003)465.
\bibitem{9} L.E. Zakharov and V. D. Shafranov, Evolution of
Equilibrium Toroidal Plasmas in Plasma Physics , MIR physics series,
Moscow (1981).
\bibitem{10} V. Arnold and B. Khesin, Topological Methods in Hydrodynamics,
Applied Mathematics Sciences 125 (1991).Imperial College Press.
\bibitem{11} S. Childress and A. Gilbert, Stretch, Twist and Fold: The Fast Dynamo (1996)(Springer).
\bibitem{12} B. Khesin, Topology Bounds Energy, in reference $1$.
\end{thebibliography}
\end{document}